\documentclass[twocolumn,showpacs,preprintnumbers,amsmath,amssymb]{revtex4}

\usepackage{graphicx}
\usepackage{dcolumn}
\usepackage{bm}

\begin{document}

\title{Scanning tunneling microscopy study of the CeTe$_{3}$ charge density wave}

\author{A. Tomic}%
\author{Zs. Rak}
\author{J. P. Veazey}
\author{S. D. Mahanti}
\author{S. H. Tessmer}
 \email{Corresponding author. E-mail: tessmer@pa.msu.edu}
 \affiliation{Department of Physics and Astronomy, Michigan State University, East Lansing, MI 48824}
\author{C. D. Malliakas}%
\author{M. G. Kanatzidis}
 \affiliation{Department of Chemistry, Northwestern University, Evanston, IL 60208}

\date{\today}

\begin{abstract}

We have studied the nature of the surface charge distribution in CeTe$_{3}$. This is a simple, cleavable, layered material with a robust one-dimensional incommensurate charge density wave (CDW). Scanning tunneling microscopy (STM) has been applied on the exposed surface of a cleaved single crystal. At 77~K, the STM images show both the atomic lattice of surface Te atoms arranged in a square net and the CDW modulations oriented at 45$^{\circ}$ with respect to the Te net. Fourier transform of the STM data shows Te square lattice peaks, and peaks related to the CDW oriented at 45$^{\circ}$ to the lattice peaks. In addition, clear peaks are present, consistent with subsurface structure and wave vector mixing effects. These data are supported by electronic structure calculations, which show that the subsurface signal most likely arises from a lattice of Ce atoms situated 2.53 ~{\AA} below the surface Te net. 

\end{abstract}

\pacs{61.44.Fw, 68.37.Ef, 71.15.Mb, 71.45.Lr, 72.15.-v, 73.20.-r, 73.20.At}

\maketitle

\section{\label{sec:intro}Introduction}

The metallic compound CeTe$_{3}$ belongs to a family of layered RETe$_{3}$ materials, where RE is a rare-earth element, which have received much attention as a model system to study incommensurate charge density waves (CDWs)~\cite{dimas;prb95, patsc;pccp02, malli;jacs127}. This class of materials features two-dimensional square-net motifs composed of Te atoms. Such square-net arrangements have been considered theoretically performing electronic band structure calculations and were found to be susceptible to CDW formation driven by Fermi surface nesting~\cite{treme;jacs87}. On the experimental side, measurements have shown rather weak coupling between the layers and large energy gaps as high as 400~meV for CeTe$_{3}$ ~\cite{dimas;cm94, ru;prb06, broue;prl04}. The CDW is well formed in CeTe$_{3}$ at room temperature, and no transition to a non-CDW state has been observed for temperatures as high as 500~K~\cite{malli;jacs128}. Although it is well established that the CDW forms in the Te net, the exact nature of the CDW in the RETe$_{3}$ family has not been resolved to date. For example, there is an ongoing debate regarding whether the CDW is uniformly incommensurate or locally commensurate within domains, with phase slips, i.e. discommensurations, occurring at the domain walls~\cite{kim;prl06, fang;prl07}.

In this paper we present scanning tunneling microscopy (STM) measurements obtained at a temperature of 77~K and theoretical calculations of the CeTe$_{3}$ surface, with the main focus on understanding the surprisingly-large subsurface contribution to the tunneling signal. This study can be compared to room temperature STM data presented in Ref.~\cite{kim;prl06}. Ref.~\cite{kim;prl06} also included Pair Distribution Function analysis of x-ray diffraction data that showed clear evidence for discommensurations; moreover, peaks in the Fourier transform of the STM images were identified as satellite structures, consistent with the discommensuration picture. The data presented here are of higher quality and are compared to a compelling wave-vector-mixing analysis, originally suggested by Fisher {\it et al.} ~\cite{fang;prl07, fisher;priv}, which does not involve discommensurations.

\begin{figure}[tbp]
\begin{center}$\,$
\includegraphics[width=3.2in,keepaspectratio=1]{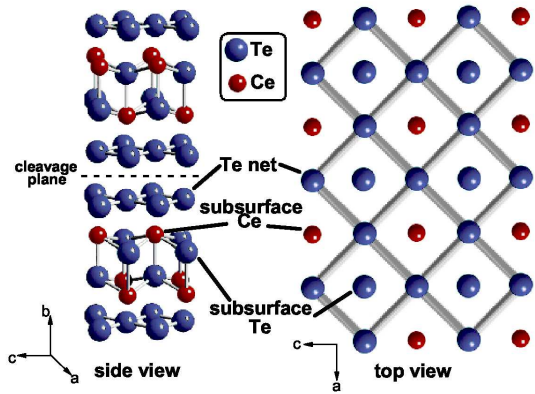}
\end{center}
\vspace{-0.2in} \caption {The average (undistorted) crystal structure of CeTe$_{3}$ consisting of corrugated CeTe slabs,
and Te layers, where Te atoms are separated by 3.1~{\AA} in a square-net. The side view shows two complete unit cells, with
the unit repeated along the b axis. The top view shows the surface Te net exposed upon cleaving the crystal. The first subsurface
layers of Ce and Te atoms are indicated. Taking into account atomic relaxation, with respect to the b-direction (perpendicular to the layers), the subsurface Ce lattice is 2.53 ~{\AA} below the surface Te net; the subsurface Te lattice is 3.44 ~{\AA} below the Te net.  
}
\protect\label{fig;CeTe3-structure}
\end{figure}

The undistorted crystal structure of CeTe$_{3}$ is shown in Fig.~\ref{fig;CeTe3-structure}. It is of NdTe$_{3}$~\cite{lin;ic65,norli;ic66} type,
weakly orthorhombic and described within the space group $Cmcm$. It is a layered structure that consists of two building blocks: double layers of [Te]$^{-}$ square-nets, and puckered ionic [Ce$_{2}^{3+}$Te$_{2}^{2-}$]$^{2+}$ double layers that are placed between the nets. The three-dimensional structure is composed of slabs of these structural motifs. The atoms within slabs are covalently bonded, while bonds between the slabs are weak
van der Waals type, allowing the crystals to cleave easily between the Te layers. Hence the exposed surface is the Te-net, ideal for STM studies of the CDW.

%
%
%
%

The existence of a unidirectional CDW in tritellurides was first reported in a transmission electron microscopy study of a series of
RETe$_{3}$ crystals by DiMasi and collaborators~\cite{dimas;prb95}. They identified superlattice reflections in the electron diffraction
pattern corresponding to a single incommensurate modulation wave vector with a magnitude of q$_{CDW}$ $\approx$ 2/7 $\times$
2$\pi$/c, where c=$\sqrt{2}a_{0}$ and $a_{0}$ is the Te-Te separation of 3.1~{\AA}. This indicated the presence of incommensurate distortions in the Te layer planes. The incommensurate superstructure was solved
within space group $C2cm(00\gamma)000$ for RE = Ce, Pr, and Nd by Malliakas and collaborators using single crystal x-ray
diffraction~\cite{malli;jacs127}. This study quantified distortions in the Te net, and revealed that a distribution of planar Te-Te distances
exists, with a minimum value of $\sim$~2.95~\AA~and a maximum value of  $\sim$~3.24~\AA .

Due to the large lattice constant along the b-direction ($\sim$~25~\AA), as shown in Fig.~\ref{fig;CeTe3-structure}, the
Brillouin zone of RETe$_{3}$ is compressed (plate-like) and slightly orthorhombic~\cite{komod;prb04}. Strong anisotropy has been observed
in their transport properties~\cite{dimas;prb95,ru;prb06}, which reflects the nearly two-dimensional
nature of the system originating from the weak hybridization between the Te layers and the RETe slabs. Electronic structure for tellurium
planes is rather simple. The electronically active valence band consists predominantly of 5$p$ orbitals of the Te atoms from the Te
planes. The only significant role in the formation of the CDW is played by the perpendicular chains of in-plane 5$p_{x}$ and 5$p_{y}$ orbitals, since the energy
of completely filled 5$p_{z}$ is pushed below the Fermi level as indicated by the first principle band structure
calculations~\cite{kikuc;jpsj98,laver;prb05}.

\section{\label{sec:exp1}STM Measurements}

\begin{figure}[tbp]
\begin{center}$\,$
\includegraphics[width=3.0in,keepaspectratio=1]{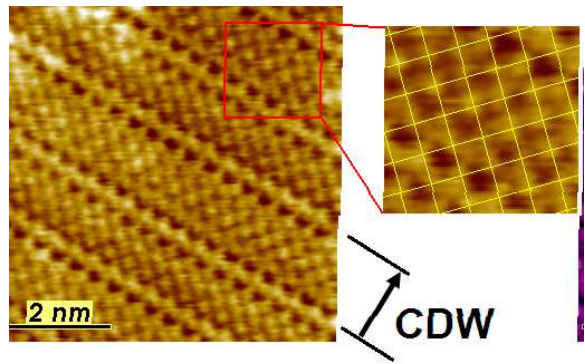}
\end{center}
\caption {A real-space STM image of the Te net obtained at 77~K, showing both Te atoms and CDW
modulations oriented at 45$^{\circ}$ to the net. The image is the average of four images that were obtained consecutively, in constant-current mode, at a at a sample bias of 100~mV and tunneling current of 0.6~nA; no image processing or filtering of the data was performed. Here we show the largest area without substantial contamination, although some is present, as seen in the upper left corner. The lines on the lower right indicate locations of high charge density due to the CDW, while the arrow marks the CDW wave-vector direction. The expanded image on the upper right includes a square grid as a guide to the eye; a Te atom is located at each intersection.}
\protect\label{fig;stm-zoom-lt}
\end{figure}

We have performed low-temperature STM topography and spectroscopy of CeTe$_{3}$ at 77~K to characterize the CDW state. CeTe$_{3}$ crystals were grown by a halide flux method, as described in Ref.~\cite{iyeir;prb03}. The crystals were carefully cleaved with adhesive tape and quickly placed in a vacuum system for subsequent STM measurements at a temperature of 77~K. A representative unfiltered STM image is shown in Fig.~\ref{fig;stm-zoom-lt}. Both the net of Te atoms and the CDW modulation are clearly visible, as indicated. The CDW modulation is oriented at 45$^{\circ}$ to the Te net.

%
%
%
%

\begin{figure}[tbp]
\begin{center}$\,$
\includegraphics[width=3.0in,keepaspectratio=1]{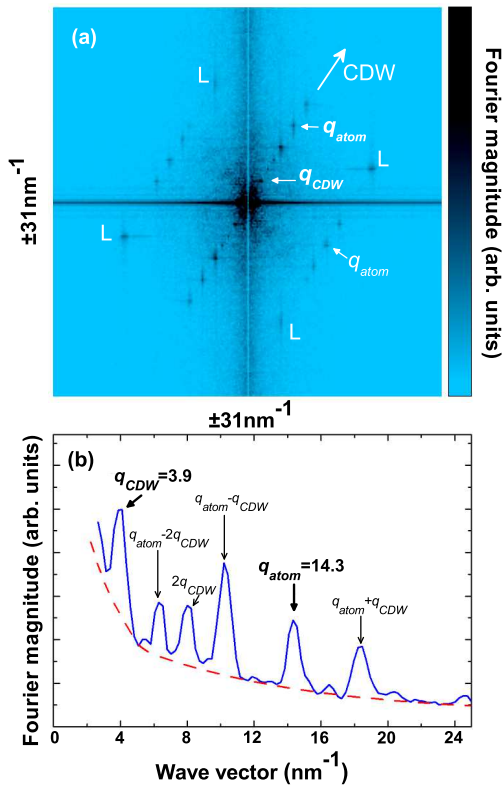}
\end{center}
\caption {\textbf{(a)} The Fourier transform of the low temperature STM data. Enhanced noise along the vertical
axis is an artifact due to the scan direction. Horizontal and vertical axes are wave vector components $k_{x}$ and $k_{y}$. The
square Te net gives rise to four distinct peaks (L). Peaks at 45$^{\circ}$ to Te net are consistent with the CDW peaks. The CDW
peak $q_{CDW}$, as well as peak $q_{atom}$, which is consistent with underlying atomic structure, are labeled. \textbf{(b)} Subset of the data from the Fourier transform along a path from the origin in the
direction of the CDW. Noise in the Fourier transform becomes significantly larger near the origin. This is due to the impurities
present in the real-space data. As the guide to the eye, the dashed line indicates the background noise.}
\protect\label{fig;ft-lt}
\end{figure}

Fig.~\ref{fig;ft-lt}~(a) shows the Fourier transform of the 77~K real-space data obtained from 16 24x24~nm images using a straightforward averaging procedure~\cite{tomic;thesis}. The Te square lattice peaks are labeled L. Peaks related to the CDW are oriented 45$^{\circ}$ clockwise to the square
lattice peaks. To examine the peaks carefully, we take a line cut along the CDW direction in Fig.~\ref{fig;ft-lt}~(b).
We expect the CDW peak to be located near $q_{CDW}$ $\approx$ 2/7 $\times$ 2$\pi$/c = 4.1~nm$^{-1}$. Indeed, we find a prominent peak at 3.9~nm$^{-1}$, which we label $q_{CDW}$ in bold.  The peak at 14.3~nm$^{-1}$, labeled $q_{atom}$, also appears in the direction perpendicular to the CDW.  Hence it is consistent with a larger square lattice oriented at 45$^{\circ}$ to the surface Te net.  As seen in Fig.~\ref{fig;CeTe3-structure}, this pattern is consistent with either the first subsurface Ce layer or the first subsurface Te layer.  A similar Fourier transform peak was observed by Fang {\it et al.}, while working with the related material TbTe$_{3}$, who also attributed it to sensitivity to the subsurface structure~\cite{fang;prl07}. Surprisingly, $q_{atom}$ and $q_{CDW}$ have roughly the same magnitude, which is approximately equal to the magnitude of the Te-net Fourier peaks, labeled L in Fig.~\ref{fig;ft-lt}~(a).  Theoretical calculations given in section~\ref{sec:theory} will address the relative magnitudes of the contributions of the tunneling current from the surface Te net and subsurface Ce and Te atoms.

\begin{figure}[tbp]
\begin{center}$\,$
\includegraphics[width=3.0in,keepaspectratio=1]{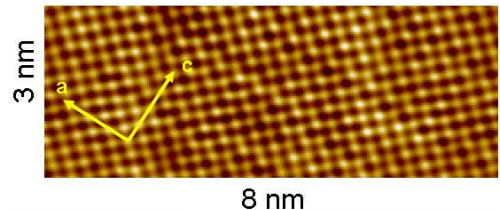}
\end{center}
\caption {STM topography of CeTe$_3$ at a sample bias of -800~mV and tunneling current of -0.05~nA, at room temperature in ambient conditions, the same tunneling conditions as Fig. 2(a) of reference~\cite{fang;prl07}.  We do not observe evidence of dimerizations, in contrast to the work of Fang {\it et al.} with the related material, TbTe$_3$. These data were filtered to reduce noise.}
\protect\label{fig;no-dimers}
\end{figure}

Because the Fourier transform lacks phase information, the square lattice corresponding to peak $q_{atom}$ could possibly be attributed to the apparent dimerizations of the surface Te net reported by Fang and coworkers while working with the related material TbTe$_3$~\cite{fang;prl07}.  However, we observe no evidence of dimers in topographic images of CeTe$_3$ when reproducing the same tunneling conditions (Fig.~\ref{fig;no-dimers}).  Moreover, we have not seen evidence for dimers in the hundreds of images we have taken of CeTe$_3$ over a broad set of bias voltages in the range of $\pm$800~mV.

%
%
%
%

%
%
%
%

\section{\label{sec:wavevectormixing}Wave vector mixing}

Returning to Fig.~\ref{fig;ft-lt}~(b), four peaks are present in addition to $q_{CDW}$ and $q_{atom}$.  In our original study, we interpreted similar structure as satellite peaks, which supported the interpretation of a discommensurated CDW~\cite{kim;prl06}. However, Fisher has suggested an alternative wave-vector-mixing explanation, which we explore here in detail~\cite{fisher;priv}.

Consider an STM signal acquired on a surface with a uniformly incommensurate CDW and lattice modulation, both of which are described by sinusoidal waves with wave vectors \(k_{CDW}\) and \(k_{atom}\) along the x-direction. The Fourier transform along this direction would ideally exhibit exactly two peaks, one at \(k_{CDW}\) and one at \(k_{atom}\). Implicit in the ideal case is the assumption that the STM signal is proportional to the sum of the two sine waves. However, in reality the signal may have a significant component that resembles the product of the two waves. In reference~\cite{tomic;thesis}, Tomic showed explicitly that this is indeed the case if an asymmetry existed in the way the CDW couples to the peaks of the atomic signal compared to the troughs. In other words, the signal will exhibit a contribution similar to the product of the two waves if the CDW signal is stronger at the atomic lattice sites and weaker at the locations between atoms, or vice versa. Algebraically, the product of two sine waves can be expressed as a sum and a difference: \(sin(k_{CDW}x)sin(k_{atom}x)=\frac{1}{2}cos(k_{CDW}x-k_{atom}x)-\frac{1}{2}cos(k_{CDW}x+k_{atom}x)\). Hence we expect this effect to give extra peaks in the Fourier transform at \(k_{CDW}+k_{atom}\) and \(k_{CDW}-k_{atom}\). More generally, Tomic showed that such a peak-trough asymmetry leads to wave-vector mixing, additional Fourier peaks at linear combinations of the two wave vectors.

With this in mind, we interpret the additional peaks in the FT of our STM data as linear combinations of \(q_{CDW}\) and \(q_{atom}\). In Fig.~\ref{fig;ft-lt}~(b), the additional peaks are labeled as \(q_{atom}-2q_{CDW}\), \(2q_{CDW}\), \(q_{atom}-q_{CDW}\) and \(q_{atom}+q_{CDW}\). The arrows mark the precise wave vector value at which we would expect each mixed peak to occur; indeed we see excellent agreement between the arrows and the actual peak locations. This analysis is similar to the analysis presented by Fang {\it et al.} to describe the TbTe$_{3}$ data~\cite{fang;prl07}.

As all of the clear Fourier peaks are accounted for without invoking discommensurations, our STM data of CeTe$_{3}$ are consistent with a uniformly incommensurate CDW. However, this analysis does not preclude the presence of discommensurations. It is possible that some of the unlabeled features such as the small peaks near 16.5~nm$^{-1}$ and 20~nm$^{-1}$ are satellite peaks indicative of discommensurations ~\cite{kim;prl06}. However these features are just above the noise level of the data.

Given that CDW and lattice modulations are typically well-described by undistorted sinusoidal waves, it is surprising that the wave-vector-mixing effect is so large. In light of the analysis by Tomic and our assertion that the lattice signal in this case corresponds to the subsurface Ce or Te, we conjecture the following. The CDW modulates the amplitude of the tunneling signal of the surface Te atoms; these are located at "trough" positions with respect to the subsurface lattice. But the CDW is expected to only weakly couple to the subsurface atoms themselves. We believe that this is the source of the peak-trough asymmetry. Unfortunately, the effect is rather subtle with respect to the real-space patterns; given the noise in the measurement, it is difficult to confirm this conjecture by examining the direct images.

\section{\label{sec:exp2}Spectroscopy}

Formation of the CDW state is expected to be accompanied by an energy gap opening up at the Fermi level. Using the point spectroscopy
mode we can probe the CDW gap at different locations of the sample surface and estimate its size. Spectra were acquired at a temperature of 77~K at various locations of the tip above the exposed Te plane on the surface of CeTe$_{3}$.

%
%
\begin{figure}[tbp]
\begin{center}$\,$
\includegraphics[width=3.0in,keepaspectratio=1]{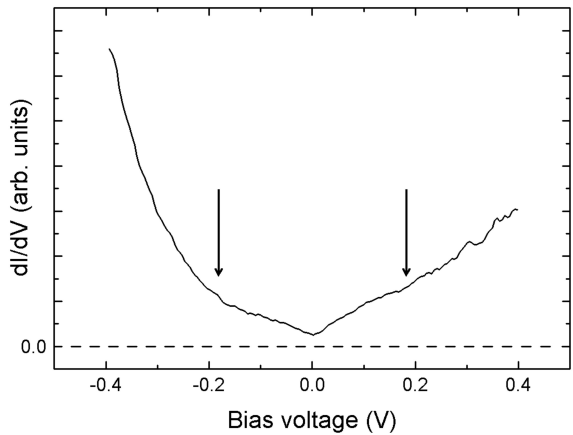}
\end{center}
\caption {Local DOS around the Fermi level for CeTe$_{3}$. The Fermi
level corresponds to zero bias voltage. The estimated CDW gap size
in CeTe$_{3}$ is about 360~meV, as indicated by vertical arrows. The
spectroscopy was carried out when the tip was located directly above
a Te atom. The effect of thermal smearing of the data is approximately 3.5~$k_{B}T$ $\approx$ 25~meV at 77~K temperature, hence it is not significant over the plotted range.}
\protect\label{fig;cete3-spect1}
\end{figure}
%
%

Fig.~\ref{fig;cete3-spect1} shows the characteristic density of states as obtained when the tip is above a Te atom. The data represent an average of 168 measurements performed consecutively. Arrows mark the approximate edges of the CDW gap, estimated to have a width of 360~meV. The magnitude and shape of the $dI$/$dV$ curves are remarkably similar to the previous TbTe$_3$ measurement by Fang and coworkers and may be compared to the ARPES results of 400~meV~\cite{broue;prl04,fang;prl07}. The local density of states has a shape that is suggestive of subgap states, characterized by the non-zero V-shaped structure within the gap.

\section{\label{sec:theory}Theoretical studies}

\subsection{\label{subsec:method}Method of Calculation}

The STM data obtained for CeTe$_3$ was simulated using electronic structure calculations within density functional theory (DFT)~\cite{hohenberg_kohn;pr64_pr65}. It is well known that the local (spin) density approximation L(S)DA fails to describe the correct ground state of a systems containing transition metal or rare-earth metal atoms. L(S)DA always puts the partially filled $d$ or $f$ bands right at the Fermi level ($E_F$), predicting metallic character with itinerant $d$ or $f$ electrons, which is obviously not correct. Strong Coulomb repulsion between localized $d$ (or $f$) electrons suppresses the charge fluctuations inherent in a metallic system. In order to describe correctly the ground state of such systems, one has to go beyond standard L(S)DA, and take into account the strong electron-electron correlations. One of the successful approaches is the L(S)DA+U method~\cite{anisi;prb91,anisi;prb93,czyzy;prb94,anisi;jphys97}, in which the localized $d$ or $f$ electrons and the delocalized $s$ and $p$ electrons are treated differently. The orbital-dependent Coulomb potential is only taken into account for localized states, while the delocalized states are treated by orbital-independent L(S)DA type potential. For CeTe$_{3}$ the LSDA+U approach was used based on full potential linearized augmented plane wave + local orbital (FPLAPW+lo) method~\cite{singh;kluver_acad94,cotte;leuven02} as implemented in the Wien2k package~\cite{blaha;austria01}. The on-site electron correlation was taken into account for the Ce $f$-states with $U_{eff}(Ce) = 6.8$~eV. For the exchange and correlation functional the local spin density approximation (LSDA)~\cite{perde;prb92} was used. The value of the  convergence parameter $RK_{max}$, which is defined as the product of the minimal atomic sphere radius $(R)$ and the largest reciprocal lattice vector $(K_{max})$
 of the plane wave basis, was chosen as $RK_{max}$ = 7. We use 2.5~a.u. for the muffin-tin radii of all Ce and Te atoms. The Brillouin zone (BZ) was sampled by a dense mesh of 625 irreducible k-points in the $k_{z} = 0$ plane. Spin-orbit (SO) interaction was included using the second variational treatment~\cite{koell;jphysc77,mcdon;jphysc80}.

\subsection{\label{subsec:stm} STM Simulation}
%
%
\begin{figure}[tbp]
\begin{center}$\,$
\includegraphics[width=2.5in,keepaspectratio=1,angle=90]{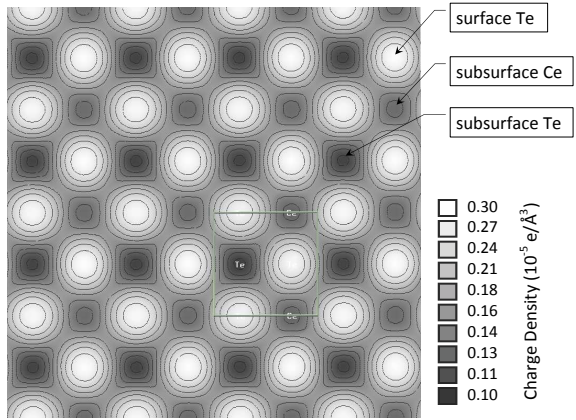}
\end{center}
\vspace{-0.2in} \caption {To simulate the STM image, we give the contour plot of the charge density projected onto a plane at 3.0~{\AA} above the surface, in the energy range of $0.0-0.1$~eV. The brighter (white) spots come from the surface Te. The subsurface Ce atoms contribute more to the charge density than the subsurface Te atoms. The Fermi level ($E_F$)corresponds to 0.0 eV. Moreover, the subsurface modulation, considering only the regions between surface Te, is of comparable magnitude to the signal from the surface Te net.}
\protect\label{fig;surf_chg_density}
\end{figure}
%
%

The CeTe$_{3}$ surface was modeled by a periodic slab geometry separated by a vacuum region of 14~{\AA}. The slabs consisted of repeating supercells, each containing two unit cells along the crystallographic long axis (b-axis for CeTe$_{3}$). The supercell was constructed using the theoretical lattice constants of the bulk CeTe$_{3}$. The thickness of the slab constructed this way is 47.5~{\AA}, which is sufficient to eliminate the spurious surface-surface interaction. Geometry optimization was performed on the surface Te square lattice and on the Ce and Te layers closest to the surface. The calculations were carried out on the high symmetry structure. Since calculations using supercell models for the incommensurate CDW are not feasible, we focus on the nature of the electronic wave functions near $E_{F}$, on the undistorted structure, with the goal of elucidating the contributions from the subsurface atoms to these wave functions.

According to the theory of tunneling between a real solid surface and model probe tip, the tunneling current in the first-order perturbation theory is given by~\cite{barde;prl61, terso;prl83}:

\begin{equation}
I = \frac{2\pi e}{\hbar}\sum_{\mu\nu}f(E_\mu)\left[1-f(E_{\nu}+eV)\right]\times\left|M_{\mu\nu}\right|^{2}\delta(E_{\mu}-E_{\nu}),
\label{eq1}
\end{equation}
where $f(E)$ is the Fermi function, $V$ is the applied voltage, $E_{\mu}$ and $E_{\nu}$ are the energies of the states $\psi_{\mu}$ and $\psi{_\nu}$, of the probe tip and the surface, respectively. $M_{\mu \nu}$ is the tunneling matrix element between $\psi_{\mu}$ and $\psi{_\nu}$. Taking the limit of small voltage (100~mV) and low temperature (77~K) and approximating the tip with a spherical wave function, the tunneling current can be written as:

\begin{equation}
\label{eq2}
I \propto \sum_{\nu}\left|\psi_{\nu}(r_{0})\right|^{2}\delta\left(E_{\nu}-E_{F}\right),
\end{equation}
where $E_{F}$ is the Fermi energy. When $E_{\nu}= E_{F}$, the tunneling current is proportional to the local density of states (LDOS) at the position of the tip ($r_0$), as given by equation~\ref{eq2}. Using the slab geometry described above we simulate a constant-height-mode STM image by calculating the charge density in an energy range of $0.0-0.1$~eV around $E_{F}$ taken at a distance of 3.0~{\AA} above the surface Te net, where 0.0 eV is the Fermi energy. Fig.~\ref{fig;surf_chg_density} shows the contour plot of this charge density. The main contribution comes from the surface Te atoms arranged in a square lattice. The subsurface Ce and Te atoms are arranged in puckered double layers in which the Ce and Te atoms form square lattices oriented at 45$^\circ$ to the surface Te net. As shown in Fig.~\ref{fig;surf_chg_density}, the subsurface Ce contributes more to the charge density than the subsurface Te, suggesting that tunneling is more likely to appear from the Ce sites than from the subsurface Te sites. This is consistent with the experimental finding that the peak q at 14.3 nm$^{-1}$ in Fig.~\ref{fig;ft-lt} corresponds to the underlying structure closest to the surface, namely to the square lattice of Ce atoms.

We have obtained similar results when we calculated the LDOS in an energy range of $\pm$0.08~eV, at distances of 2.5~{\AA} and 3.0~{\AA} above the surface Te net.

\subsection{\label{subsec:electronic_struct} Electronic Structure}

%
%
\begin{figure}[tbp]
\begin{center}$\,$
\includegraphics[width=3.0in,keepaspectratio=1]{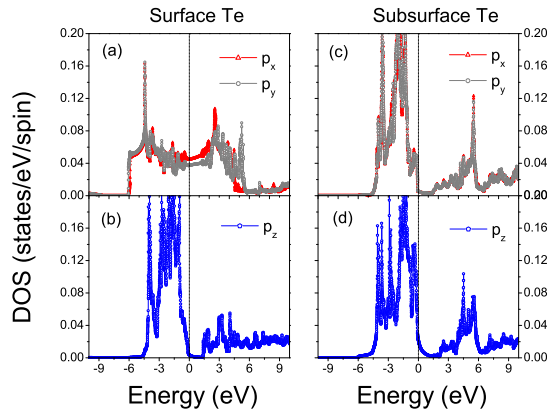}
\end{center}
\vspace{-0.2in} \caption {The only significant contribution to the DOS near $E_{F}$ comes from the surface Te $p_{x}$ and $p_{y}$ states.(a) Partial DOS associated with the $p_{x}$ and $p_{y}$ orbitals of the surface Te atoms. The flat DOS near $E_{F}$ is characteristic to the CDW systems. (b) The $p_{z}$ orbital of the surface Te is occupied and located mostly below $E_{F}$. (c),(d) The $p$-states of the subsurface Te atoms are occupied, because the trivalent Ce provides two electrons within the Ce-Te slab.}
\protect\label{fig;Te_dos}
\end{figure}
%
%

%
%
\begin{figure}[tbp]
\begin{center}$\,$
\includegraphics[width=3.25in,keepaspectratio=1]{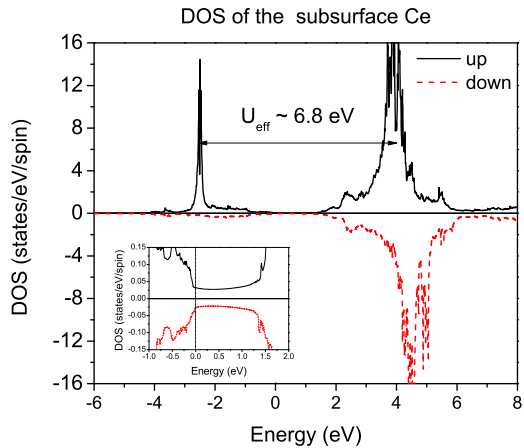}
\end{center}
\vspace{-0.2in} \caption {Spin-polarized, partial DOS associated with the subsurface Ce atom. The peak at -2.5~eV in the spin-up DOS corresponds to one occupied f-level while the peaks between 3.0-5.0~eV (spin-up and spin-down) represent the 13 unoccupied f-states. The inset shows that the DOS near $E_{F}$ is nonzero.}
\protect\label{fig;Ce_dos}
\end{figure}
%
%

Figure~\ref{fig;Te_dos} shows the calculated partial density of states (PDOS) associated with the $p$-orbitals of the surface and subsurface Te atoms. Since the spin-up and spin-down DOS associated with the Te atoms are identical, we only show the spin-up channel. As pointed out by DiMasi et al.~\cite{dimas;cm94,diams;prb52} in RETe$_{3}$ compounds the rare earth is trivalent, it provides two electrons for bonding to the puckered double-layer of RE-Te and one electron to the Te square net. Therefore we expect the $p$-states of the subsurface Te atoms (which are part of the RE-Te slabs) to be completely occupied, while the $p$-states surface Te atoms (which form the square net) should be partially filled. As shown in Fig.~\ref{fig;Te_dos}, our electronic structure calculation agrees with this observation and with the earlier theoretical results \cite{kikuc;jpsj98,laver;prb05}: the only significant contribution to the DOS at $E_{F}$ comes from the $p_{x}$ and $p_{y}$ orbitals of the surface Te. The partial DOS associated with these two orbitals (Fig.~\ref{fig;Te_dos}~(a)) are rather flat at $E_{F}$, which is a characteristic of the CDW systems in the high symmetry phase. The $p_{z}$ orbital of the surface Te atom (Fig.~\ref{fig;Te_dos}~(b)) and the the $p$-states of the subsurface Te atoms (Fig.~\ref{fig;Te_dos}~(c),~(d)) are nearly fully occupied and located mostly below $E_{F}$.

In order to understand why the subsurface Ce atoms contribute so much to the STM charge density calculation, we give the spin polarized DOS associated with the subsurface Ce atom in Fig.~\ref{fig;Ce_dos}. The spin-up states are represented in the positive (upper) region, while the spin-down states are shown in the negative (lower) region of the graph. The main contribution to the DOS comes from the Ce $f$-states: the narrow, sharp peak in the spin-up DOS located at $\sim$~2.5~eV below $E_{F}$ represents the only one occupied Ce $f$-level. The broader peaks located between 3.0 eV and 5.0 eV above $E_{F}$, in both the spin-up and spin-down DOS, come from the rest of the 13 empty Ce $f$-states (6 spin-up and 7 spin-down states). The splitting between the occupied and empty states is approximately equal to the chosen value of the Coulomb repulsion within the $f$-shell $U_{eff}(Ce) = 6.8$~eV. The inset of Fig.~\ref{fig;Ce_dos} gives the DOS associated with the Ce orbitals near $E_{F}$. There is a small but finite contribution coming from both the hybridization of $f$-states with Te $p$-bands and Ce $d$ states. These states will contribute  to tunneling measurements above the Ce sites.

Now let us discuss the tunneling results in the light of our theoretical calculations. As discussed in sec.~\ref{subsec:stm}, the STM tunneling current is proportional to the LDOS at the position of the tip. In reality however there is a matrix element between the states from which the tunneling occurs (host) and the tip state. Assuming that the tip state can be approximated by a smooth $s$-like function, the symmetry of the host state will determine the strength of the tunneling current. For example, for the surface Te, $p_{x}$ and $p_{y}$ states that make dominant contribution to the DOS near $E_{F}$ will have zero (very small) contribution to the tunneling current. As discussed above, the intensity of the FT of the STM signal as a function of q along the  direction of the CDW wave-vector (which is rotated by 45$^\circ$ with respect to the direction of the square lattice peaks) gives peaks at $q_{CDW}$ = 3.9~nm$^{-1}$ and at $q_{atom}$ = 14.3~nm$^{-1}$, the latter is ascribed to the subsurface Ce or Te lattice. From our PDOS calculations we find that for the subsurface Te, the $p_{x}$ and $p_{y}$ orbitals contribute $\sim$~0.004~stats/eV/spin near EF, whereas the $p_{z}$ orbital contributes $\sim$~0.02~states/eV/spin. Thus both due to symmetry and small PDOS, the $p_{x}$ and $p_{y}$ states will not contribute to the tunneling current, only the $p_{z}$ orbital will contribute. The PDOS associated with the Ce orbitals are $\sim$~0.03~states/eV/spin comparable to the subsurface  Te $p_{z}$. However the lattice of Ce atoms is a distance of 2.53 ~{\AA} from the surface Te net, whereas the subsurface Te lattice is a distance of 3.44 ~{\AA}. Hence the Ce atoms are nearly 1 ~{\AA} closer to the STM tip; it is therefore likely that the tunneling current above Ce sites is larger than that above the subsurface Te sites. Furthermore, since the surface Te $p_{z}$ contribution to the DOS near $E_{F}$ is rather small ($\sim$~0.004~stats/eV/spin) our charge density analysis suggests that the tunneling currents above the Ce sites and the surface Te net sites are comparable. This is the origin of the surprisingly high sensitivity to the subsurface Ce, consistent with both the experiment, (Fig.~\ref{fig;ft-lt}) and theoretical simulation (Fig.~\ref{fig;surf_chg_density}). This result is significant because it may be possible to probe the nature of Ce $f$ states through careful tunneling measurements.

\section{\label{sec:summary}Summary}

STM constant-current-mode images and spectroscopy measurements of CeTe$_3$ were acquired at 77~K. The effects of the CDW were clearly resolved in both the images and the spectroscopy curves. In addition to the CDW, the images show the expected signal from the surface net of Te atoms and a large contribution from subsurface structure of approximately the same magnitude. Contrary to a study by Fang {\it et al.} ~\cite{fang;prl07}, no evidence of dimerization of the surface Te net was observed.

Fourier transform analysis of the STM images showed two principal peaks along the direction of the charge density wave, one from the CDW at $q_{CDW}$ =3.9~nm$^{-1}$ and the other from a subsurface lattice at $q_{atom}$ = 14.3~nm$^{-1}$. In addition to these peaks four others were observed which we show are well described by a wave-vector-mixing effect of the principle wave vectors. We believe the high degree of wave vector mixing is caused essentially by the fact that the CDW, which exists in the surface layer, distorts preferentially the troughs between the subsurface atoms. With regard to the uniformity of the CDW, our present study does not provide support for discommensurations, in contrast to the interpretation of our earlier STM work ~\cite{kim;prl06}. However, the data presented here do not rule out discommensurations and the issue remains an open question.  

To better understand the nature of the tunneling signal and the large contribution from subsurface atoms, we theoretically considered the symmetry of the tunneling matrix elements and performed density function theory calculations. We concluded that the dominant contribution to the subsurface signal is from a lattice of Ce atoms 2.53 ~{\AA} below the surface Te net. Moreover a simulated STM image constructed from our calculations confirms that the modulation of the signal arising from this lattice is comparable in magnitude to the signal from the surface Te net.

\begin{acknowledgments}
We gratefully acknowledge helpful comments and advice from S.J.L. Billinge, H.J. Kim and E.S. Bozin. Theoretical calculations were performed at the High
Performance Computing Center at Michigan State University.  This work was supported by the National Science Foundation, Grant Nos. DMR-0305461, DMR-0703940 and DMR-0801855.

\end{acknowledgments}

\end{document}